# PANIC: the new panoramic NIR camera for Calar Alto

Harald Baumeister[a*], Matthias Alter[a], M. Concepción Cárdenas Vázquez[b], Matilde Fernandez[b], Josef Fried[a], Jens Helmling[c], Armin Huber[a], Jose-Miguel Ibáñez Mengual[b], Julio F. Rodríguez Gómez[b], Werner Laun[a], Rainer Lenzen[a], Ulrich Mall[a], Vianak Naranjo[a], Jose-Ricardo Ramos[a], Ralf-Rainer Rohloff[a], Antonio García Segura[b], Clemens Storz[a], Marcos Ubierna[b], Karl Wagner[a]

[a]Max-Planck-Institut für Astronomie, Königstuhl 17, 69117 Heidelberg, Germany
[b]Instituto de Astrofisica de Andalucía, Cno. Bajo de Huertor, 50, 18008 Granada, Spain
[c]Centro Astronomico Hispano Aleman, C/ Jesús Durbán Remón, 2-2, 04004 Almeria, Spain

## ABSTRACT

PANIC is a wide-field NIR camera, which is currently under development for the Calar Alto observatory (CAHA) in Spain. It uses a mosaic of four Hawaii-2RG detectors and covers the spectral range from 0.8-2.5 µm (z to K-band). The field-of-view is 30x30 arcmin. This instrument can be used at the 2.2m telescope (0.45arcsec/pixel, 0.5x0.5 degree FOV) and at the 3.5m telescope (0.23arcsec/pixel, 0.25x0.25 degree FOV).
The operating temperature is about 77K, achieved by liquid Nitrogen cooling. The cryogenic optics has three flat folding mirrors with diameters up to 282 mm and nine lenses with diameters between 130 mm and 255 mm. A compact filter unit can carry up to 19 filters distributed over four filter wheels. Narrow band (1%) filters can be used.
The instrument has a diameter of 1.1 m and it is about 1 m long. The weight limit of 400 kg at the 2.2m telescope requires a light-weight cryostat design. The aluminium vacuum vessel and radiation shield have wall thicknesses of only 6 mm and 3 mm respectively.

## 1. INTRODUCTION

The greatest strength of the Calar Alto observatory is its up-to date instrumentation. Whereas the 3.5m telescope is well equipped with modern instruments, the 2.2m telescope is lacking modern instrumentation; an exception is the lucky imager Astralux, but this is a special purpose instrument for a limited range of applications.

A survey on instrumentation wanted for Calar Alto at MPIA and IAA came independently to the same result, i.e. a NIR (0.9 to 2.5 µm) imager for the 2.2m telescope. It is obvious that an instrument with a single 2x2 k detector is not state of the art, so a mosaic of 4 detectors was envisaged. Science applications would obviously be wide field (>1 square degree) imaging and surveys, but a collection of ideas at IAA and MPIA showed that there are many very interesting applications also for single pointed observations. PANIC is not the first instrument of its kind, but the proposed science cases show that interesting science can be done with it, even if it is not unique. Therefore the Calar Alto Instrumentation Committee recommended to build such an instrument. In October 2006 the project PANIC (**Pa**noramic **N**ear **I**nfrared Camera for **C**alar Alto) was started. It is a joint project between IAA and MPIA.

Basic requirements which were clear from the start are: detector size 4096x4096 pixels, spectral range YJHK minimum, image scale 0.45 arcsec/pixel. The pixel scale is obviously a matter of applications; however, it turned out that PANIC can be used at both the 2.2m and the 3.5m telescope if the cold pupil stop can be changed. The image scale on the 3.5m telescope will be 0.23 arcsec/pixel, so PANIC can be used in a low spatial resolution wide field mode on the 2.2m telescope and for high spatial resolution observations on the 3.5m telescope.

---

[*] baumeister@mpia.de; phone +49-6221-528316

The limits for size and weight at the 2.2m telescope require a folded design with a light-weight cryostat (see below). Since the standard telescope guiding unit would vignette the field, PANIC will make use of the on-chip guiding capability of the Teledyne Hawaii-2RG detectors.

## 2. INSTRUMENT OVERVIEW

PANIC has a 2x2 array of Hawaii2-RG NIR detectors with a gap of 3 mm between the single detectors. These detectors are mounted on a common plate, which is part of the detector unit, described in section 3.

The instrument has a diameter of 1100 mm and a height of 1000 mm. Its total mass is only 390 kg, including liquid nitrogen. The mass is limited by the 2.2 m telescope to 400 kg. Therefore a challenging light-weight design of the cryostat, telescope adapter and the optics is required.

The cryostat has two liquid nitrogen vessels: A big vessel for cooling the radiation shield and the optics and a smaller one for cooling the FPA. The extremely light-weighted cold bench has a diameter of 1050 mm, a thickness of 40 mm and a weight of 33 kg, which is only about 40 % of the mass of a solid plate with the same dimensions. The fused silica entrance window of the cryostat is 330 mm in diameter and it is 20 mm thick.

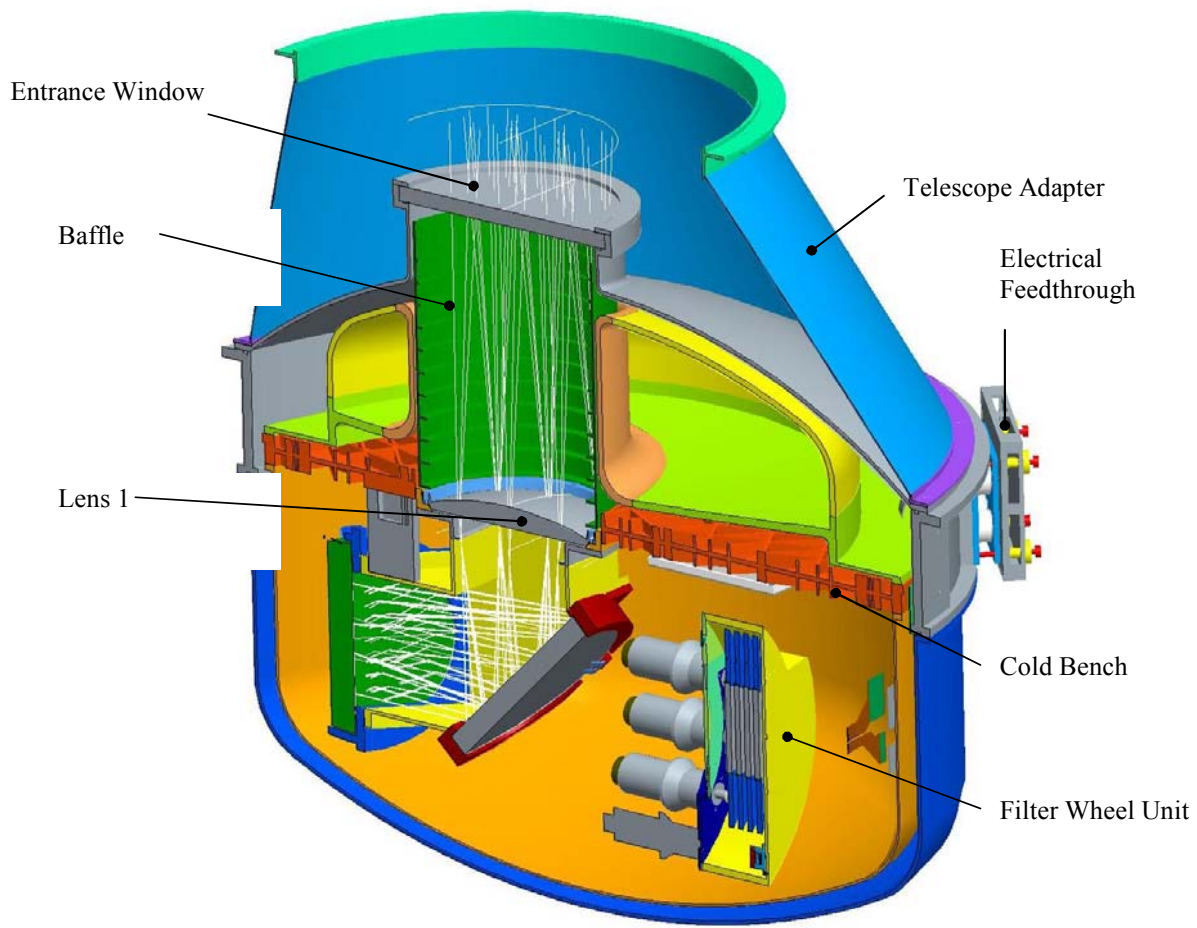

Fig. 1. Vertical section through PANIC.

The cold optics uses nine lenses and three folding mirrors, arranged in two main groups. The lenses have diameters between 130 mm and 255 mm. Several optical materials, such as fused silica, $CaF_2$, $BaF_2$ and S-FTM16 from Ohara are used. The folding mirrors are made from fused silica. They have diameters between 229mm and 282 mm. For details on the optical components and their mounts see section 5.2.

The filters of this instrument are kept in four filter wheels. Each wheel has six positions, thereby allowing for 19 filters (1 dark and 4 open) altogether. The filters are made from fused silica or N-BK7 (to be determined) and they have diameters of 125 mm.

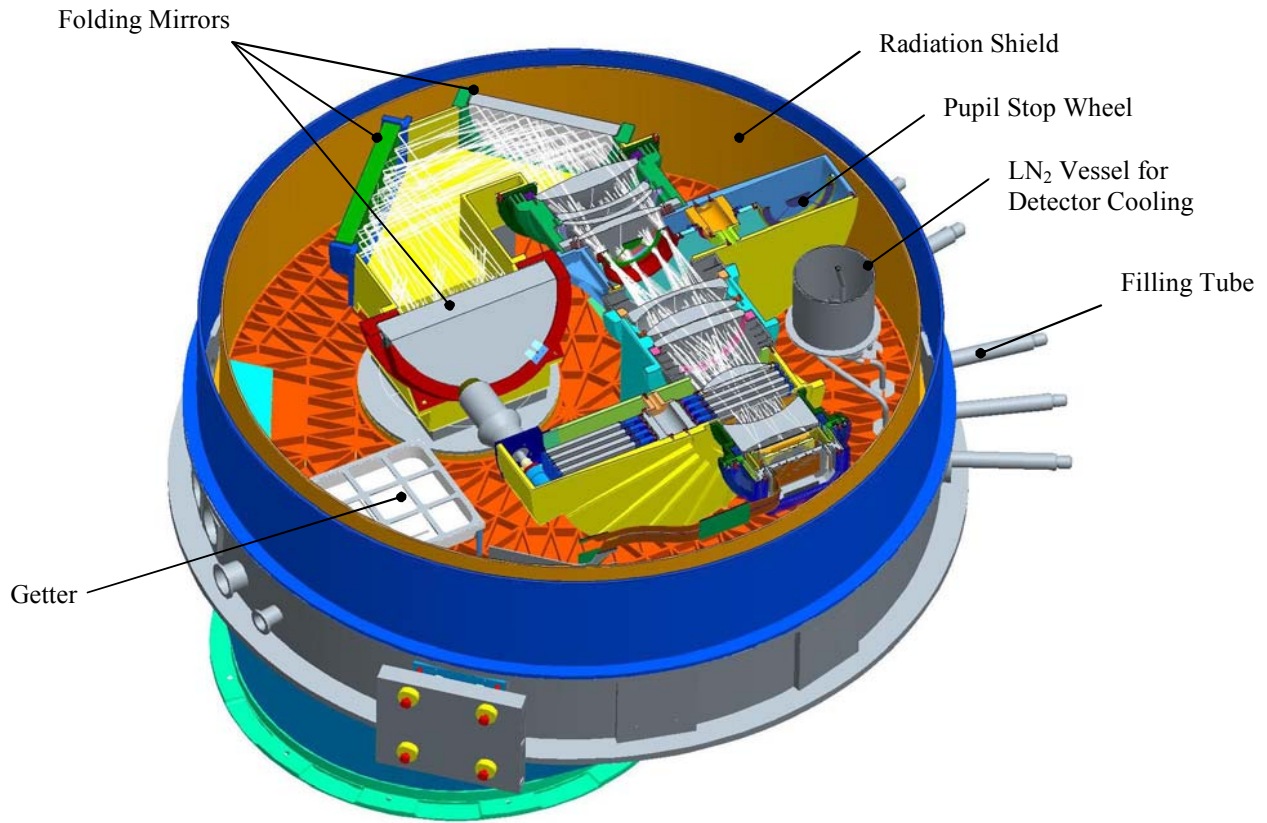

Fig. 2. Section through PANIC showing the cold bench and the cold optics.

## 3. NIR DETECTORS

The science detector for PANIC is a mosaic of four 2k x 2k 18 μm pixel HAWAII-2RG, MBE (Molecular Beam Epitaxy) grown HgCdTe devices. Because of their anti-reflection coating and substrate-removed characteristics, these SCAs (Sensor Chip Assembly) have high QE ($\geq 75$ %) over their entire operating spectral range [5]. PANIC covers z- to K-bands (from 0.8 μm to 2.5 μm). These detectors allow on-chip guiding which is important since the standard telescope guiding units would vignette the field. Thus a guider can be realized without additional hardware.

The individual SCAs are mounted into a mosaic module which holds the new 3-legged H2RG package devices used for PANIC. The mosaic, designed by GL Scientific, can be simplified into 3 parts: a top cover, a molybdenum structure, and an aluminum base plate. While the top cover carries a light shield and serves as housing for the detectors, the molybdenum structure acts as a common assembly plate that integrates the four HAWAII-2RG arrays into one single thermal and vibration-stable configuration, allowing precision alignment and physical flatness between all 4 detectors.

The surface of each detector, as well as over the whole focal plane array (4 detectors), is specified to a flatness of 80 μm peak to peak (± 40μm from the best fit plane). Fig. 3 shows an example of the SCAs mounted into the mosaic module. The optical gap between the individual detectors corresponds to 3 mm, approximately 75 arcsec at PANIC's pixel scale at the 2.2 m telescope (40 arcsec at the 3.5 m).

The bottom part of the module consists of an aluminum base plate that holds not only the molybdenum structure, but also the electrical feedthroughs for the detectors and the connectors for the two independent control loops in charge of their thermal stability. In the case of PANIC, it also serves as the mechanical interface for integrating the module into the cold optics.

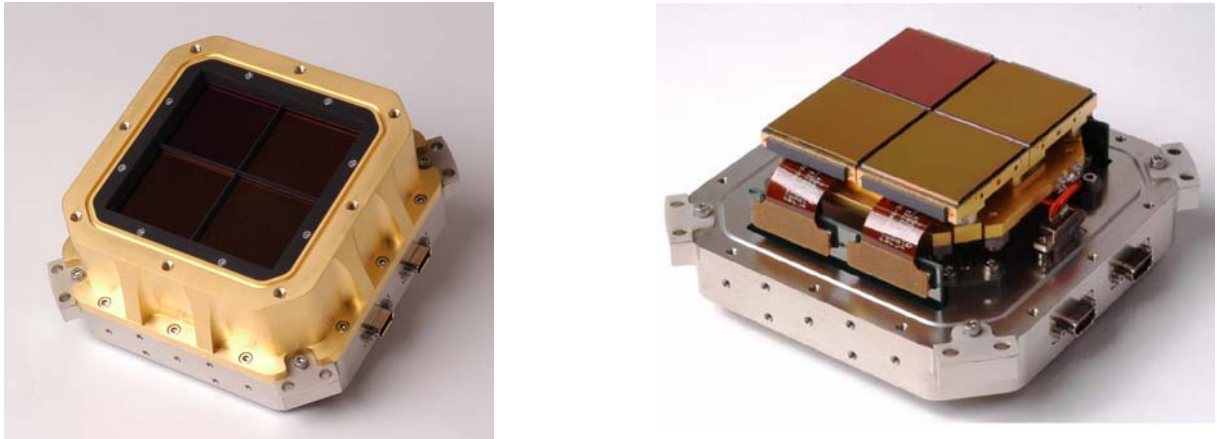

Fig. 3: Example of the mosaic module (left) and four H2RG's mounted into it (right).
Courtesy of Teledyne Scientific and Imaging, LLC.

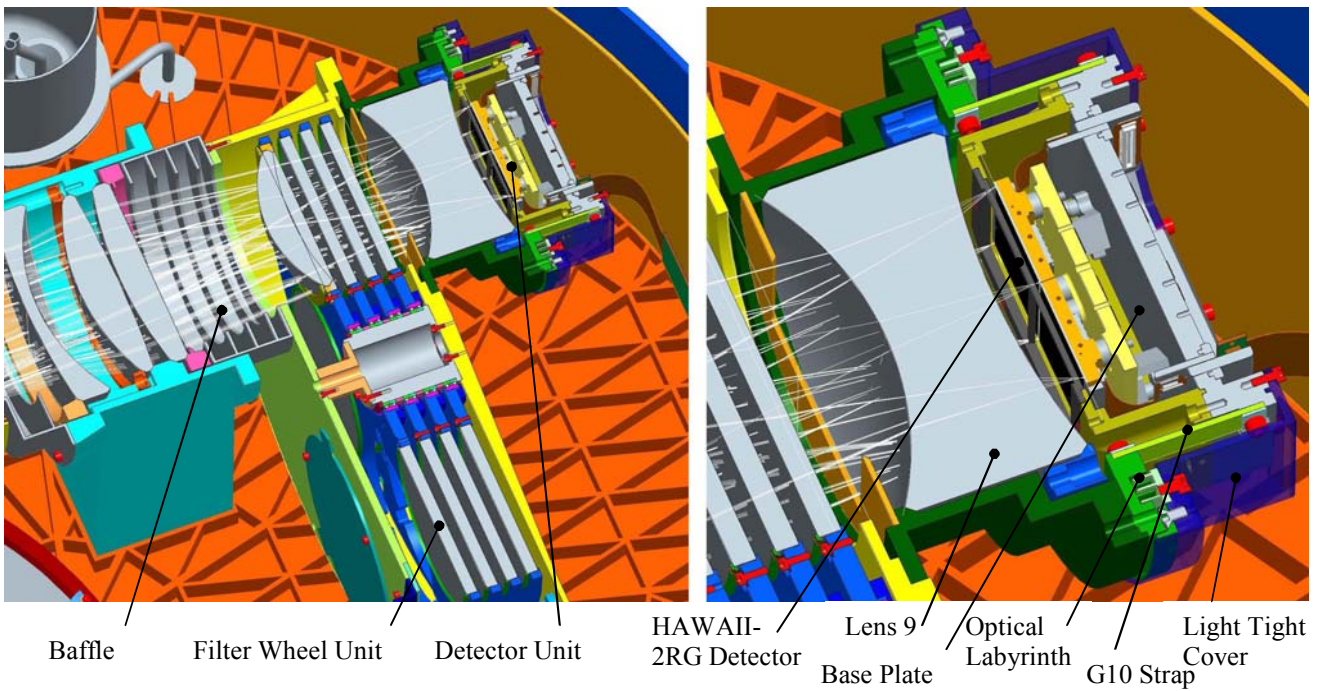

Baffle   Filter Wheel Unit   Detector Unit   HAWAII-2RG Detector   Lens 9   Optical Labyrinth   Light Tight Cover
                                             Base Plate                    G10 Strap

Fig. 4: The mosaic is mounted behind the optics unit, mechanically attached to the optics wheel housing.

Each of the detectors is electrically divided into 32 vertical stripes that correspond to the output video channels. The channels can be read out in parallel with pixel rates up to 5 MHz, allowing the implementation of a fast guiding window using a sub-frame of the detector, while at the same time reading the full array at a lower speed. The read out noise for low speed readout (100 KHz) is specified to 20 e- for CDS (Correlated Double Sampling) [6].

## 4. CRYOSTAT

The instrument is built for a wavelength range from 0.8 to 2.5 µm. This requires a temperature of the cold optics below 100 K. The focal plane array is optimized for a working temperature of about 77 K. All this can easily be achieved by a $LN_2$ bath cryostat. A big vessel will cool the complete optical bench, together with the optics and the shielding. A second, smaller vessel will cool the FPA only. This concept was chosen because of thermal drifts due to the movement of the instrument and the changing filling level. The changes could result in thermal gradients and differences of about 10 K. This is not a problem for the optics but it is unacceptable for the FPAs.

The optics and FPAs are mounted to a stable optical bench. This bench is bolted by spacers to a central ring of the vacuum can. This central ring supports all the feedthroughs for electronics, nitrogen and vacuum. It should not be necessary to take the bench and ring apart. Connected to this ring will be the telescope adapter, which has an unusual shape because the central instrument axis and the optical axis of the telescope have a shift of about 170 mm.

If the instrument is pointing to zenith, the large nitrogen vessel is located above the bench and the optics below. The optics is covered by a shield and this is enclosed in a convex vacuum can. There will be no feedthrough or any other connection at this shield on this part of the vacuum can. This allows easy opening of the cryostat and free access to the cold optics.

The large nitrogen vessel above the optical bench needs to have a big tube because the light path goes right through the vessel.

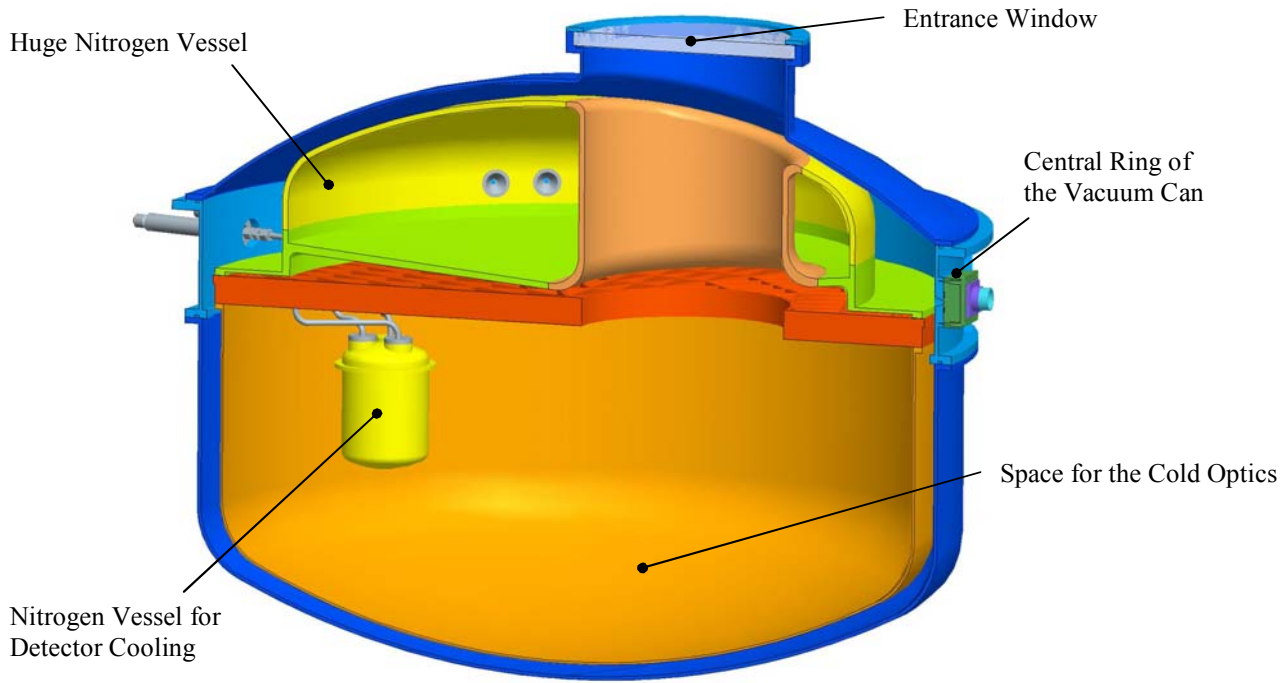

Fig. 5: Section through the cryostat of PANIC.

From the first concept weight reduction had a high priority. We have used convex ends to reduce the mass of the vacuum can and the $LN_2$ vessels. This allows thinner walls. All vessels are made from aluminum. FEM has shown that the vessels are stiff enough for the pressure difference. In addition, we have to conform with the European Pressure Equipment Directive (97/23/EC). This will be double checked and certified by an external company.

To reduce the size of the nitrogen vessel, we have to reduce the heat input to reach a hold time of more than 24 hours. Therefore, we will use MLI (multi layer insulation) to reduce the major contribution of heat input, which is the heat radiation from room temperature. This gives also the advantage of independence from the outside temperature and reduced thermal gradients.

The bench temperature will only passively be cooled by $LN_2$. The resulting temperature will slightly vary in a range of about 10 K. This is within the thermal requirements for the cold optics. The influence from the outside world on the detectors is small because everything is enclosed in the cold optics volume. The detector will then be equipped with heater and temperature sensor, stabilized by a controller. The complete FPA is thermally isolated from the optical bench.

## 5. CRYOGENIC OPTICS AND MECHANICS

The cold optics is grouped into two main assemblies, which are mounted separately to the bench close to their respective centre of gravity. The first optics assembly contains lens 1 and the three folding mirrors. Lenses 2 to 9, the pupil stop exchanger, the filter wheel unit, and the detector unit are arranged in the second main assembly. Table 2 shows the group arrangement of the cold optics. The optical design is described in detail in [7].

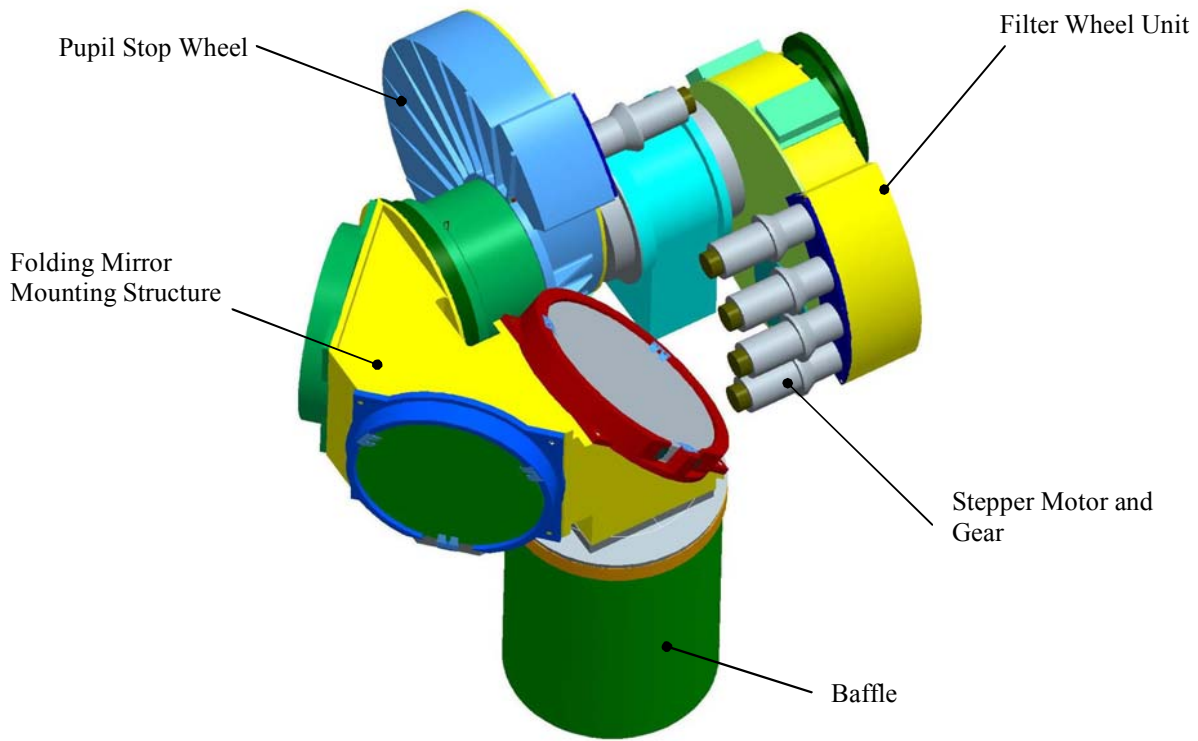

Fig. 6: Cold Optics with filter wheel unit, pupil stop exchanger, baffles and detector unit.

| Optical element | L1 | M1 | M2 | M3 | L2 | L3 | L4 | L5 | Cold stop | L6 | L7 | L8 | L9 | FPA |
|---|---|---|---|---|---|---|---|---|---|---|---|---|---|---|
| **Groups** | LM 1 | mirror structure | | | LM 2a | | | | LM 2b | | LM 3 | | LM 4 | |
| | optics mount 1 | | | | | | | | | optics mount 2 | | | | |
| | complete optics | | | | | | | | | | | | | |

Table 1: Optical groups of PANIC including the focal plane array. (LM … lens mount, L … lens, M … mirror)

## 5.1 Cryogenic lenses and their mounts

The cryogenic lens mounts use the same principle as OMEGA2000 [1] and PYRAMIR [2][3]. The most difficult task of the lens mount design is to make sure that the lenses survive cooling and at the same time achieve the tight tolerances required by the optical design. If the lenses were mounted in a conventional way, e.g. with a threaded retainer ring, the different thermal expansion properties of the materials used might lead to severe damage during cooling. Therefore, we employ a mounting method that uses chamfers at both the lenses and the mount parts. In this case, we chose a chamfer angle of 40° for both outer edges of each lens, the lens mount and the retainer ring. The lenses sit in the conical surfaces of the mount. The retainer rings keep the lenses in this position by the forces of eight disk spring packages each. Temperature changes result in diameter changes of the parts. These changes lead to an axial displacement of the lenses and retainer rings because the parts can slide on the chamfer surfaces relatively to each other, assuming that the chamfers are manufactured very precisely and that friction can be neglected. The mounting principle is described in detail in [1].

|        | Material     | Diameter [mm] |
|--------|--------------|---------------|
| **L1** | Fused silica | 255 |
| **M1** | Fused silica | 282 |
| **M2** | Fused silica | 256 |
| **M3** | Fused silica | 229 |
| **L2** | $CaF_2$      | 170 |
| **L3** | S-FTM16      | 152 |
| **L4** | Fused silica | 154 |
| **L5** | $BaF_2$      | 142 |
| **L6** | S-FTM16      | 160 |
| **L7** | $BaF_2$      | 143 |
| **L8** | Fused silica | 150 |
| **Filter** | Fused silica | 125 |
| **L9** | Fused silica | 130 |

Table 2: Cryogenic optical components of PANIC

| Material | Thermal expansion (293 K --> 77K) in % |
|----------|----------------------------------------|
| Fused silica | +0.001 |
| CaF2 | -0.284 |
| BaF2 | -0.306 |
| S-FTM16 | -0.167 |
| AlMg4.5Mn (Al5083) | -0.386 |

Table 3: Thermal expansion rate from room temperature to 77 K

## 5.2 Cryogenic mirrors and their mounts

Panic uses three cryogenic folding mirrors, which are all mounted in the same way by kinematic mounts. The mirrors are fixed in the axial direction by double spring clips, as shown in Fig. 7. The radial degrees-of-freedom are restricted by two tangential flat surfaces and a pair of spring clips. These spring clips push the mirrors against the flat surfaces. All mirror units are mounted to a single structure, which is shown in Fig. 6.

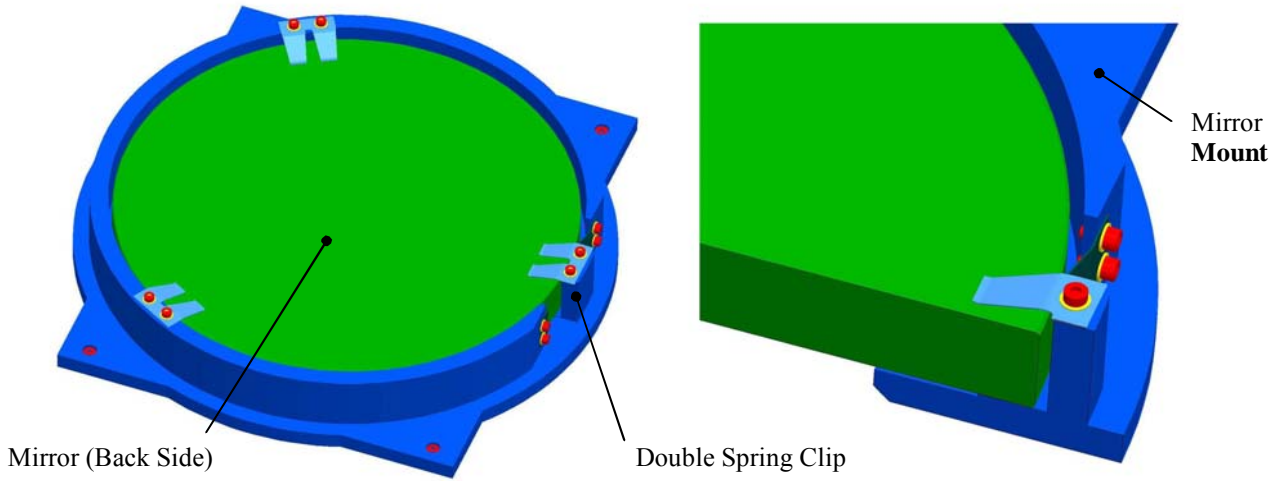

Fig. 7: Kinematic mount of one of the flat mirrors (left: total view, right: detail of section view).

### 5.3 Cold pupil stop wheel

The PANIC instrument is designed to be mounted to both the 2.2m telescope and the 3.5m telescope. Each of the telescopes requires a different size and position of the pupil stop mask. To allow quick exchange, both masks are mounted to a wheel which is driven by a cryogenic stepper motor by Phytron and a Harmonic Drive gear. Two position switches are used to detect which of the masks is in place. The drive concept is the same as for the filter wheels (see section 5.4). For easy spare part management, the same motor and gear is used for both the filter wheel unit and the pupil stop wheel.

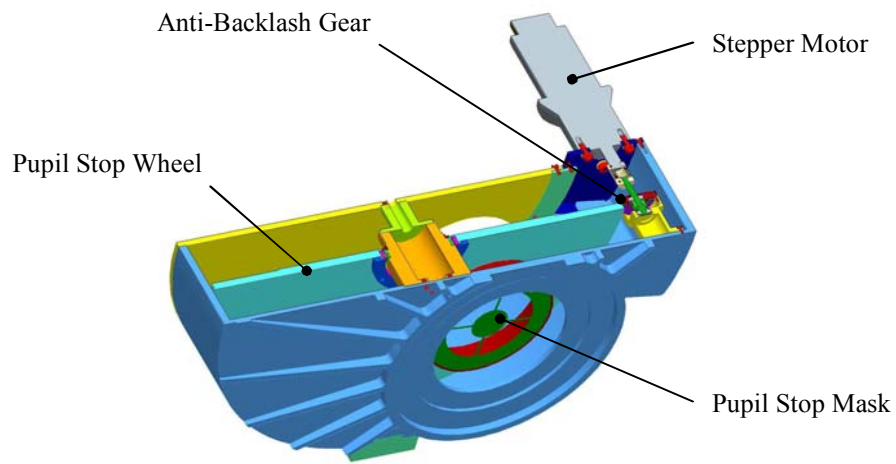

Fig. 8: Section view of the cryogenic pupil stop wheel

### 5.4 Filter wheel unit

The filter wheel unit contains four wheels with 6 positions each. One position of each wheel has to stay empty, giving 20 positions, which can be used for filters, a black disk or pupil re-imager lenses (one for each telescope). Each wheel has a pre-loaded double row ball bearing from ADR, coated with dry lubrication. Pre-loading the bearing gives a high running precision and at the same time we have a much better heat flow for the cooling of the wheel compared to a single row bearing. As a result, no additional cooling mechanism is necessary. The wheels have a toothing at their outer diameters and they are driven by Phytron cryogenic stepper motors which are equipped with modified Harmonic Drive gears and resolvers from LTN Servotechnik. These resolvers are modified for use at cryogenic temperatures. They give feedback on the motor position with a precision of 12 arcmin. Since about 12 motor revolutions are needed for one turn of the filter wheel, we need to count the motor steps from a zero position (mechanical reference switch) to know the filter wheel position. Since the space for the wheels is very limited, it is not possible mount the resolvers directly to the wheels.

A preloaded anti-backlash spur gear is used between the Harmonic Drive gears and the wheels. The spur gear – wheel combination and the Harmonic Drive together have a transmission ratio of 680:1. The stepper motor needs 200 steps for one revolution. Therefore, a filter wheel is turned by 9.5 arcsec by one motor step. The wheel positioning accuracy is dominated by the transmission accuracy of the Harmonic Drive. This is about 1 arcmin, corresponding to an angle of 4.4 arcsec at the wheel. Each filter is held by three copper-beryllium spring clips.

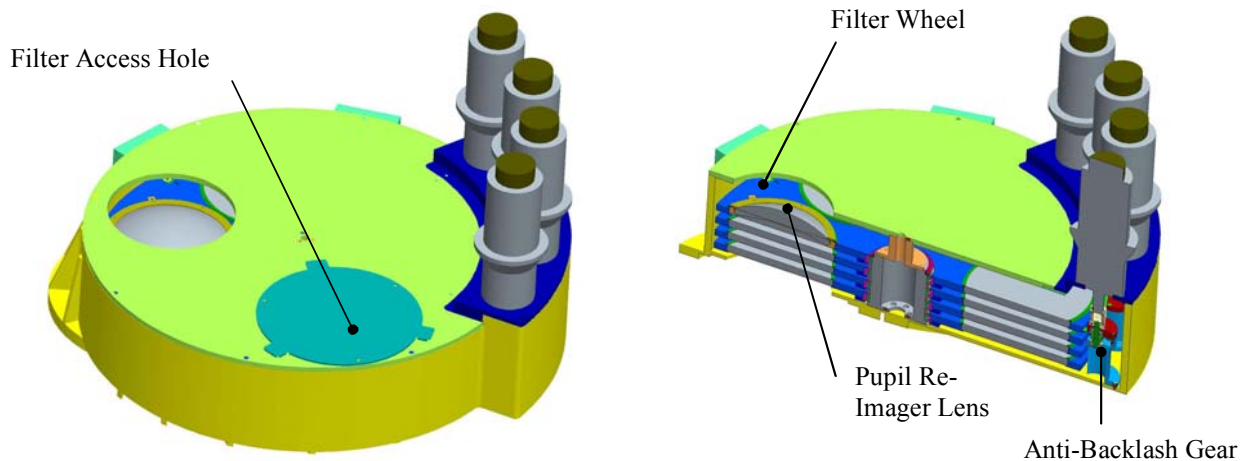

Fig. 9: Filter wheel unit with pupil re-imager lens.

## 6. CURRENT STATUS AND OUTLOOK

The project passed the PDR in November 2007. The optical design is almost finished and it is currently being iterated with potential lens manufacturers. The optics FDR will take place in July 2008 and mechanics FDR in late 2008. The detectors are ordered. They will be delivered in spring 2009. First light on the telescope is planned for early 2010.